# Mega electrorheological phenomena in graphene nano–gels

Purbarun Dhar *[,a,b], Ajay Katiyar [b,c], Arvind Pattamatta [b] and Sarit K Das [a,b]

[a] Department of Mechanical Engineering, Indian Institute of Technology Ropar, Rupnagar–140001, India

[b] Department of Mechanical Engineering, Indian Institute of Technology Madras, Chennai–600036, India

[c] Research and Innovation Centre (DRDO), IIT Madras Research Park, Chennai–600113, India

## Abstract

Unprecedentedly massive electrorheology (ER) has been reported for dilute graphene nanoflakes based ER fluids that have been engineered as novel, readily synthesizable polymeric gels. Polyethylene glycol (PEG 400) based graphene gels have been synthesized and very high ER response (~ 25,000 % enhancement in viscosity under influence of electric field) has been observed for low concentration systems (~ 2 wt. %). The gels overcome several drawbacks innate to ER fluids. The gels exhibit long term stability, high graphene packing ratio which ensures very high ER response and the microstructure of the gels ensure that fibrillation of the graphene nanoflakes under field is undisturbed by thermal fluctuations, further leading to mega ER. The gels exhibit large yield stress handling caliber with yield stress observed as high as ~ 13 kPa at 2 wt. % graphene. Detailed investigations on the effects of graphene concentration, electric field strength, imposed shear resistance and transients of electric field actuation on the ER response of the gels have been performed. The gels show nearly negligible hysteresis with respect to electro viscous effects which speak of the stability and responsiveness required in transient situations. In-depth analyses with explanations have been provided for the observations and effects, such as sheet over-crowding induced loss of structural integrity at high concentrations and inter flake lubrication/ slip induced augmented ER response. The present gels show great promise as potential ER gels for variant smart applications, such as damping and vibration control in active aircraft and vehicle structures, micro-nanoscale electro-actuation, in suspensions or braking/ clutching in high performance specialized vehicular systems, etc.

*Corresponding author: E-mail: purbarun@iitrpr.ac.in
Phone: 0–1881–24–2173
**_Note:_** *The work has been done in part during the lead author's previous affiliation at IIT Madras and in part during the present affiliation at IIT Ropar.*

# 1. Introduction

The phenomena of electroeheology (ER) [1–3] has been an area of intriguing research wihtin the academic commnunity with a strong interdisciplinary flavor. The increase in viscosity [4] by orders of magnitude of such a colloidal system and its associated control by the application of an external electric field has been subject to extensive research. In general, such fluids consist of an insulating fluid [5–7], mostly dielectric oils or organic liquids, with dielectric particles [8–11], mostly of micron scale, dispersed in it to form a stable suspension. When an electric field is applied against such a colloid, the particles align themselves along the electric field lines to form fibrils or chained structures [12–15]. This leads to transition from fluidic to a semi–solid state, which enables the fluid to resist shear deformation. The solid like character imparted to the ER fluid also ensures that the fluid exhibits finite yield stress values [16, 17] under the influence of electric field, a phenomena which is not observed in fluids. However, micron scale particles often lead to high erosion and corrosion as well as sedimentation over time. Accordingly, nanoscale particles are preferred in the present era, however, such particles are prone to thermal fluctuations and consequently, large solid like transition is difficult to obtain [18]. In many cases, the particles are coated with dielectric shells so as to increse their relative permittivity in order to counter the effects of thermal fluctuation. In certain cases, such engineered particles lead to very gigantic yield stress capabilities under the influence of electric field [19, 20].

In recent times, another possible method to counter the thermal fluctuations in nanostructures has been the use of suitable carbon based nanostructures. Since such nanostrucutres are high aspect ratio and/or non–spherical strucutres, these are weak Brownian entities and hence the fibrils or chains are not prone to thermal disturbances. Also, as such systems have low densities, the suspensions are in general stable over long time periods. However, as the relative permittivities of such nano carbon systems are not high, the nanostructures require propoer functionalization in order to obtain appreciable ER effect. The most common material is this category has been carbon nanotubes (CNTs) [21, 22], however, graphene oxide and graphene (G) [23–25] have also been reported. As of present reports, graphene based ER systems show very low values of electroviscous response (increase in viscosity with increasing electric field at a particular shear rate) as well a low values of yield stress. Since carbon based systems have low dielectric constants, high packing density is desired for the ER fluid in order to obtain very strong fibrils. However, given the large population of delocalized pi electrons in G, it is ensured that the fibril alignment along the electric field lines will be prompt and strong.

If proper packing of the G flakes can be obtained, it will lead to formation of very strong fibrils and the ER effect will be very large. When used in conjunction with liquids, the packing density is low and hence the fibrils are not strong enough to impart high ER response. Also dilute suspensions do not exhibit such rheological phenomenon since proper structuring or fibrillations are not possible at such low viscosities owing to Brownian fluctuations. Furthermore, chemical synthesis of large quantities of G without compromising its quality is exceedingly difficult and so very concentrated systems are difficult to obtain in liquid systems. In the present article, the ER response of G has been exploited using a novel technique. In order to obtain high ER from G at low concentrations, polyethylene glycol (PEG 400) based novel graphene nano gels (GNGs) have been synthesized in order to obtain the required high viscosity, closely packed fibrillation at low G concentrations. ER is a characteristic of concentrated suspensions, with the particle loading often in the range of ~ 10 to 30 vol. %. However, given the structural nature of the present gel samples, low concentrations are able to provide very high ER response. Exhaustive ER studies, such as electroviscous response of the gels at varying concentrations, electroviscous response at variant electric field transients and electric field effects on yield stress have been reported. Enhancement of viscosity by 3–4 orders of magnitude as well as yield stresses greater than 10 kPa have been observed for very low G concentrations of ~ 2 wt. %. The high enhancements have been explained based on a merger of the two prominent theories of ER, the interfacial tension thoery and the electrostatic theories. The present gels show tremendous potential as high response mega ER fluids for various smart applications.

## 2. Materials and methodologies
### 2.1. Synthesis and characterization of graphene

The nano–G samples utilized in the present report have been synthesized by chemical exfoliation of graphite by modified Hummer's method and further rendered hydrophilic in accordance to reported protocol [26]. The synthesized samples have been characterized by Raman spectroscopy and high resolution Scanning Electron Microscopy to ensure that G has been obtained. The size distribution of G flakes has been characterized using Dynamic Light

Scattering (DLS) technique. The Raman spectra, SEM image and DLS spectra for two representative synthesized samples, namely G1 and G2, have been illustrated in Figure 1. Analysis of the Raman spectrum (Fig. 1 (b)) reveals the presence of sharp, high intensity peaks at ~ 1350 cm$^{-1}$ and ~ 1560 cm$^{-1}$, which are the characteristic D and G bands of graphene systems [27]. The G band is a manifestation of the planar stretching of the sp$^2$ hybridized carbon atoms in the graphene flakes whereas the D band arises due to the surface defects and wrinkles inherent to the exfoliation process. In addition, the broadly distributed peaks at ~ 2800 cm$^{-1}$ (termed the 2D band) are also characteristic of such systems and conclusively indicate the presence of G. The ratios of the intensities of 2D and G bands have been utilized to estimate the number of layers in the synthesized G samples. In the present system, this ratio has been observed to be in the vicinity of 0.3-0.5, which implies that the population of G flakes are on average 3-10 sheets thick [28].

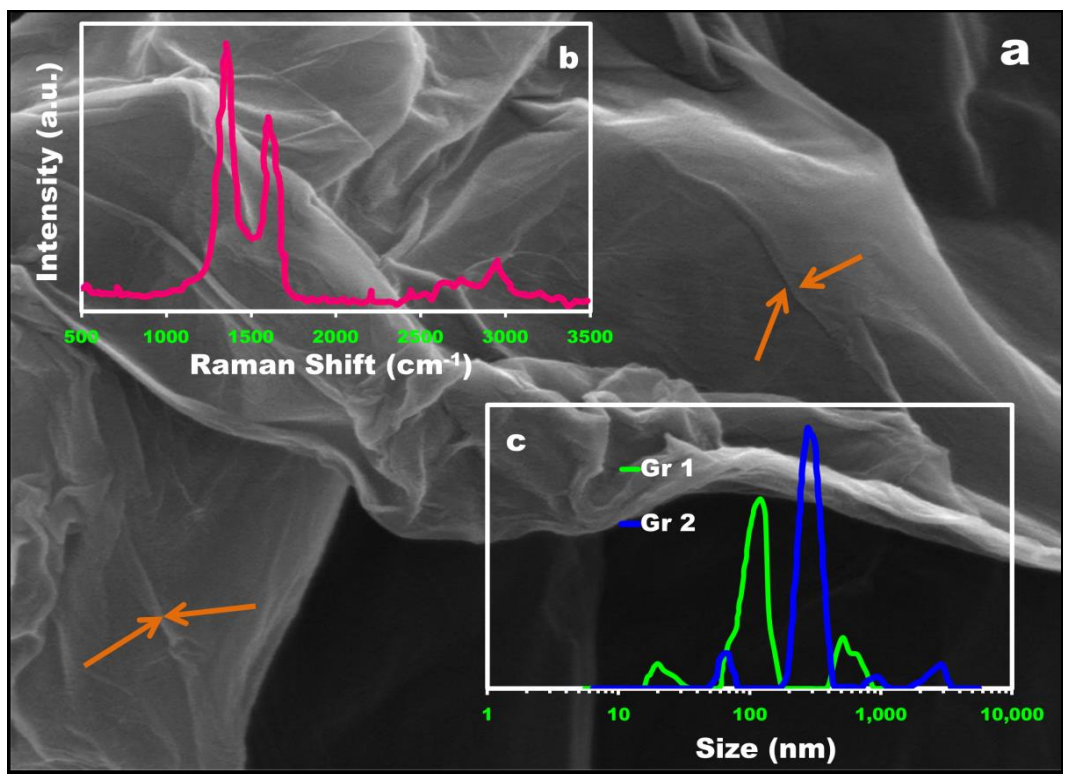

**Figure 1**: Characterization of the graphene sample utilized in the present case. (a) SEM image of the nano flakes (b) Raman spectra of the G sample (c) DLS spectra of two representative samples.

The G sample has also been characterized though HRSEM imaging which provide conclusive evidence in favor of the formation of G structures based on the surface imperfections observed. The imperfections and wrinkles appear due to stretching of the $sp^2$ hybrid layer of carbon atoms, leading to nanoscale faults and crests, as indicated by arrows in Fig. 1(a). The strong oxidizers used in the chemical protocol exfoliate the Gr stacks to form G. Due to non–uniform localized gradients and reaction kinetics, the exfoliation is seldom uniform spatially and this also leads to presence of voluminous amounts of surface wrinkles and sheet defects, features in general absent on Gr flakes. The observable surface wrinkling in SEM images in fact confirms the presence of exfoliated G structures [26]. The flake size distributions for two representative G populations, namely G1 and G2, have been obtained using DLS analysis and have been illustrated in Fig. 1(c). The DLS analysis reveals that major fraction of the flakes of sample G1 lie around the peak size of ~ 100 nm, with minor populations around 25 nm and 30 nm, whereas in case of G2, the major population is distributed around ~ 400 nm, with minor populations around 70 nm, 800 nm and 2000 nm. Therefore, the total population of G flake sizes lies within the meso–nanoscale regime.

**2.2. Synthesis of graphene gels**

The nanogels or pastes (the words gels and pastes have been used interchangeably in the present context, since based on the concentration of graphene, the texture visually emulates either a gel or a paste form) have been synthesized and reported for the first time, as an exhaustive survey of literature reveals. The gels have been synthesized utilizing DI water and polyethylene glycol of atomic weight 400 Da (PEG–400). PEG-400 has been utilized specifically since it is a readily available polymer, among one of the most utilized, highly biocompatible, non-toxic, and water soluble low molecular weight polymers. Furthermore, its thermophysical and dielectric characteristics are readily available as standardized data.

Initially, PEG–400 is mixed with DI water in the ratio 1:1 (by volume) and raw nanocolloids are prepared by dispersing G in it and ultrasonicating the same for 2–3 hours, depending on the concentration of G. The concentration of G added is evaluated as wt. % of the PEG-400 in the water–PEG solution only. The colloid (illustrated in Figure 2 (a)) is heated to ~ 75–80 $^{o}$C in a flask with a refluxing arrangement attached (boiling of the colloid is restricted). Reflux heating is done for ~ 1.5 hours to obtain an intermediate gel with slimy texture. The refluxing unit is then removed and the gel is transferred to a wide mouthed beaker and heated at a reduced temperature of 50 $^{o}$C. The wide cross section of the beaker helps in more uniform evaporation of the remaining water from the slimy intermediate. The heating is continued for 10–15 minutes with constant visual inspection such that the slime does not completely dry out to form solid films. The moment at which the vapor issuing from the gel reduces drastically, the heating is discontinued. At this point, the water content of the gel reduces to trace amounts, but just enough to retain the gel form and remains so over a long shelf life (over years with no visible separation of the trace water). The graphene nano gel (GNG) is then allowed to cool to room temperature and collected (Figure 2 (b)).

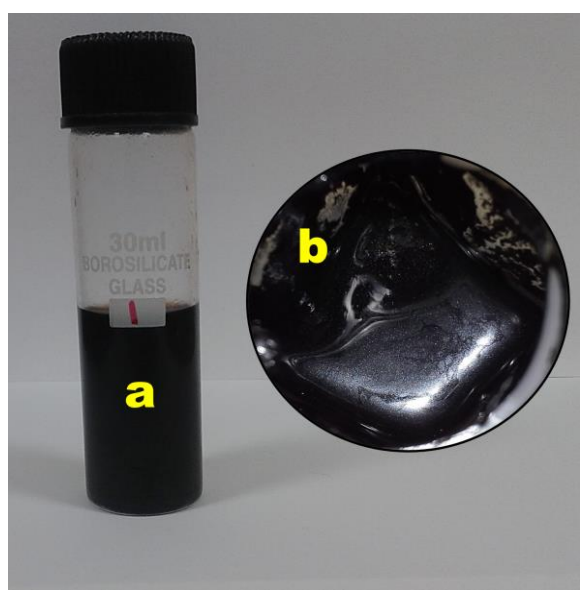

**Figure 2**: Illustration of (a) the preliminary nano graphene–water–PEG colloid and (b) graphene nano–gel post synthesis.

Utmost care is taken not to heat the gel beyond the point where the water content reduces drastically and the vapor issuance disappears completely. Heating beyond this point leads to loss of the trace amounts of water too, leading to complete drying of the gel. It has been observed from repeated trials that it is the trace water that preserves the texture and form of the gel for indefinite time. Excess water content also leads to fluidic texture of the gel and therefore care is taken during preparation through thorough monitoring. Four GNG samples, *viz.* G1, G2, G0 and G3, consisting 0.5 wt. %, 1 wt. %, 1.5 wt. % and 2 wt. % G were synthesized for evaluating electrorheological characteristics. The concentration of G in the GNGs is restricted to 2 wt. % of the PEG due to material constraints. It has been observed repeatedly during various synthesis trials that in general, samples with concentration over and above ~ 2 wt. %. fail to form gels, with odds against gel formation around 4.5 out of 5. Instead, the PEG–water–G system when refluxed, leads to caramelization of the polymer, leading to formation of a brownish semi-fluid which no further leads to gelation. Given the high thermal conductivity and heat capacity of G, presence of G nanoflakes in the PEG-water system leads to enhanced absorption of heat during refluxing. The enhanced localized heating due to presence of the G nanoflakes leads to change in the physico-chemical properties of the PEG, leading to caramelization. As a result, the polymer chains are either convoluted or broken and thus the change in microstructure induces inability to form the required gel phase.

In order to study the ER responses on the GNGs, the electrorheology module of a rheometer system (Physica MCR 301, Anton Paar, Germany) has been utilized. The module employs a parallel plate configuration wherein the sample to be tested is sandwiched between the flat stator and the rotor heads. The rotor and stator plates are connected to the terminals of the power supply unit, which can provide an HV input of up to 12 kV, and depending on the gap within which the sample is sandwiched, the corresponding electric field intensity can be deduced. Depending on the requirement, the field can be increased as a function of time, with the minimum possible increment of ~ 4 kV/s. The maximum plausible electric field intensity that can be applied depends on the dielectric breakdown strength of the sample, with the onset of breakdown detectable from a sudden rise in current in the power supply display. In the present case, the GNGs exhibit breakdown voltages in the vicinity of ~ 6–7 kV/mm and accordingly the maximum field has been restricted to 5 kV/mm.

## 3. Results and discussions
### 3.1. Response of graphene colloids to electric fields

In order to understand the mechanism behind ER response of a population of G nanoflakes, the most important phenomenon that requires to be observed is the ability and propensity of chain formation or fibrillation by the flakes under electric fields. ER response arises due to the enhanced viscous resistance imparted by the fluid due to transition from a liquid state to a partial solid like state. This transition is caused by the stacking or structuring of the constituent particles (in present case the G nanoflakes) which leads to formation of chained structures or fibrils within the fluid medium, thus imparting enhanced resistance to deformation. The process of fibrillation is, in general, difficult to quantify or model mathematically given the complex and uncertain nature of the system [13, 15]. The propensity of chain formation depends on several factors, viz. shape and size of the dispersed particles, the dielectric properties of the particles and fluid, fluid properties such as viscosity and density and the nature, strength and distribution of the electric field lines, just to name a few of the major parameters [6, 11]. However, despite the inherent difficulty in modeling the phenomenon, fibrillation can be explained qualitatively based on experimental visual observations. The fibrillation characteristics exhibit by the G nanoflakes have been visually captured utilizing microscopy (employing a ferrograph with external provisions for applying electric field across a fluid sample) and the same have been illustrated in Figure 3. High degree of polarization of the delocalized dense π electron cloud system in G nanoflakes caused by the external field, leading to electrostatic interaction based structuring and self–assembly along the field lines can be attributed to explain the fibrillation in G systems. This is somewhat similar, yet different in implications as far as electrorheology is concerned (discussed later), to G oxide systems.

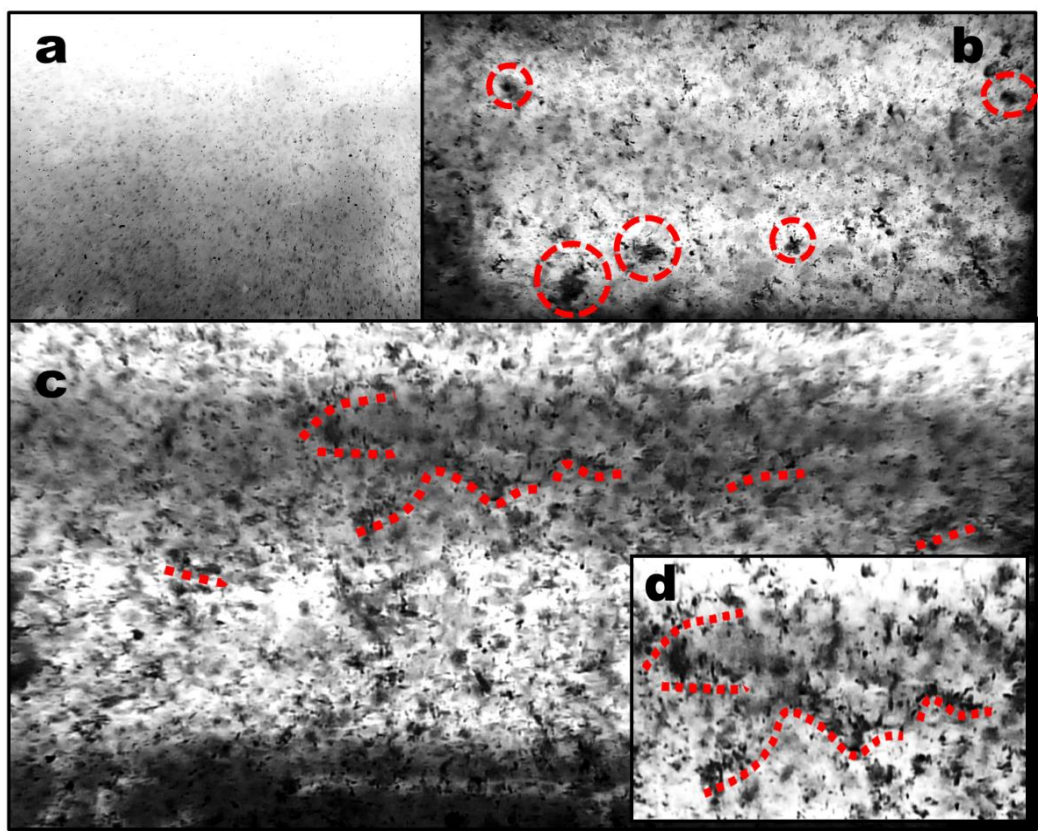

**Figure 3:** Fibrillation or chain forming propensity of G flakes under the influence of externally applied electric fields. The pictures have been obtained utilizing ferrography technique which involves obtaining microscopy images of particulate matter suspended in fluids. **(a)** The GNF in absence of field **(b)** the GNF in presence of an electric field of intensity of 0.1 kV/mm and localized grouping/ clustering of the G flakes (illustrated by circles) **(c)** the GNF in presence of an electric field of intensity of 0.2 kV/mm and localized fibrillation is observed (illustrated by dotted lines for guide to the eye) **(d)** zoomed view of a chained structure at 0.2 kV/mm field intensity. Inception of clustering and tendency towards subsequent fibrillation/ chain formation therefore occurs at very low field intensities.

For observing fibrillation capabilities, a very dilute (0.01 wt. %) aqueous dispersion of G is synthesized by sonication. Sonication has been restricted to a few minutes so as not to disperse the whole flake aggregate population to their nanoscale sizes, which would otherwise make it difficult to observe the colloid by optical microscopy. A dilute system is preferred since the backlight utilized for capturing the microscopy images can easily be transmitted across the translucent suspension, and thereby fibrillation characteristics can be observed. In order to apply an electric field, two flat aluminum sheets spaced 2 mm from each other are kept immersed in the dispersion and connected to a digitally controlled AC power supply system (AC has been employed as a DC system will induce directional electrophoretic drift and electrolysis at the metal-water interface instead of the required fibrillation). The region between the facing plates is focused using the ferrograph microscope and a snapshot of the dispersion (obtained using a digital microscope camera) in absence of field obtained has been presented in Figure 3 (a). An electric field of 0.1 kV/mm strength is applied between the two electrodes and the consequent microscopy image has been presented in Figure 3 (b). As observable, the low field intensity is potent enough to induce miniscule localized aggregations of the G flakes, among which some further aggregate to form larger conglomerates (depicted in Figure 3 (b) by dotted circles). This is evident of the fact that the highly conducting G nanoflakes align themselves in space in accordance to the localized field line distribution and form concentrated clusters, some of which grow in size due to excess agglomeration. Complete fibrillation is not possible under the influence of very low field intensities since the thermal randomness of the nanoflakes is high enough to overcome stacking of the flakes along the field lines.

Further, the electric field intensity is increased to 0.2 kV/mm and nascent stages of fibrillation are clearly visible in the microscopy image in Figure 3 (c). A visible evidence of structuring is seen within the whole colloid with the presence of small fibrils in the formative stages. For ease of comprehension, the more distinct nascent fibrils have been provided with parallel dotted lines. Figure 3 (d) illustrates a zoomed in view of one of the more prominent evidences of fibrillation. Although the aggregated flakes are able to form short fibrils at this field intensity by overcoming the momentum induced by thermal fluctuations, the interflake separation is large due to the dilute nature of the dispersion and the weak magnitude of the applied field. However, it can be observed that the inherent clustering response is higher than G oxide or other G based systems [23, 24] as nascent fibers are seem to coalesce even at low fields. It is clearly evident from Figure 3 (d) that the micro-clusters formed at 0.1 kV/mm have reassembled themselves along the field lines (directed horizontally across) and are crowding to form microfibrils. The fact that such nanoflakes can form minute agglomerates under the influence of small electric field strengths which further assemble to form nascent fibrils cements the fact that at higher fields it possible for the flakes to form strong and long fibrils. In the present case, the physical form of the GNGs, which ensures a viscous, close packed system, enhances the possibility of chain formation by eliminating thermal fluctuations and severely reducing interflake separation. Thereby, the observations and nature of the gels reveal that it is possible to obtain strong ER characteristics from such gels.

### 3.2. Electroviscous response

The ER effect in GNGs can be established from the electroviscous effects that it exhibits and hence will be the first factor of discussion in the present article. Electroviscous effect in the context of ER may be defined as the change in the viscosity as a function of applied field intensity and is often studied at different shear rates to understand the phenomenon comprehensively [5, 7]. The effect has been experimentally studied for the GNGs by applying constant shear loading while increasing the electric field intensity and recording the viscous resistance offered by the gels. The electroviscous response of the samples G0 and G3 have been illustrated in Figure 4 (a) and Figure 5 (a) respectively. The two samples have been highlighted for their electroviscous abilities since their high G content provides the highest magnitudes in enhancement of viscosity. The enhancement of viscosity obtained at 4.8 kV/mm for the samples G0 and G3 at different shear loading have been illustrated in Figure 4

(b) and 5 (b) respectively. The enormity of electroviscous effect observed, to the extent of enhancement of viscosity (at 0.001 s$^{-1}$ shear) by ~ 20000 % and ~ 4500 % for G3 and G0 respectively, are rarely obtained in ER fluids at low concentrations. Such occurrences are only reported for colloids exhibiting giant ER phenomena [19] and that too is obtained at relatively higher concentrations. Further, this effect is much pronounced than those exhibit by similar G oxide or graphite systems. The mechanism governing such enhancements can be explained based on a fusion of two major theories of ER, *viz.* the interfacial tension or water bridge theory [1, 5] and the electrostatic and/or fibrillation theory [14, 29].

The water bridge theory is valid for systems containing three components as it assumes the ER fluid as a three phase system. The dispersed phase is considered to contain the third phase (which is another liquid, e.g. water) immiscible with the primary phase liquid (e.g. oil or in this case the PEG) coated as interfacial nanofilms on the particulate surface. In the event of absence of an electric field, the third phase is strongly attached and stably adheres to the surface of the dispersed phase. In presence of an electric field, the molecules of the third phase is driven to one side of the surface of the particles by the phenomena of electrokinesis and this leads to binding the adjacent particles together by interfacial tension to form chains. As a result, the ER fluid behaves like a pseudo-solid due to change in its microstructure. Since the reflux procedure does not lead to complete removal of water from the system, it can be safely assumed that a trace percentage of water remains behind and is distributed uniformly within the GNGs. On the contrary, the electrostatic theory considers the ER fluid as a two phase system with dielectric particles forming fibrils aligned with an applied electric field. Often ER fluids are synthesized utilizing a conducting solid phase physically or chemically coated with an insulating layer and then dispersed in insulating fluid media. From microstructure considerations, it can be theorized that the present GNGs possess the dual nature of both types of ER fluids. While the system definitely contains trace amounts of water molecules which would be enough to form the interfacial films over the G nanoflakes, they cannot form bridges by interfacial tension since water and PEG are highly miscible. However, the interfacial film acts as a suitable dielectric layer around the highly conductive G nanoflakes, thereby creating an electrostatic ER fluid working on fibrillation principle. Intuitively, the transition from fluid to solid via fibrillation in the present gels can be argued to be fuelled by a hybrid mechanism of the water bridge and the electrostatic theories.

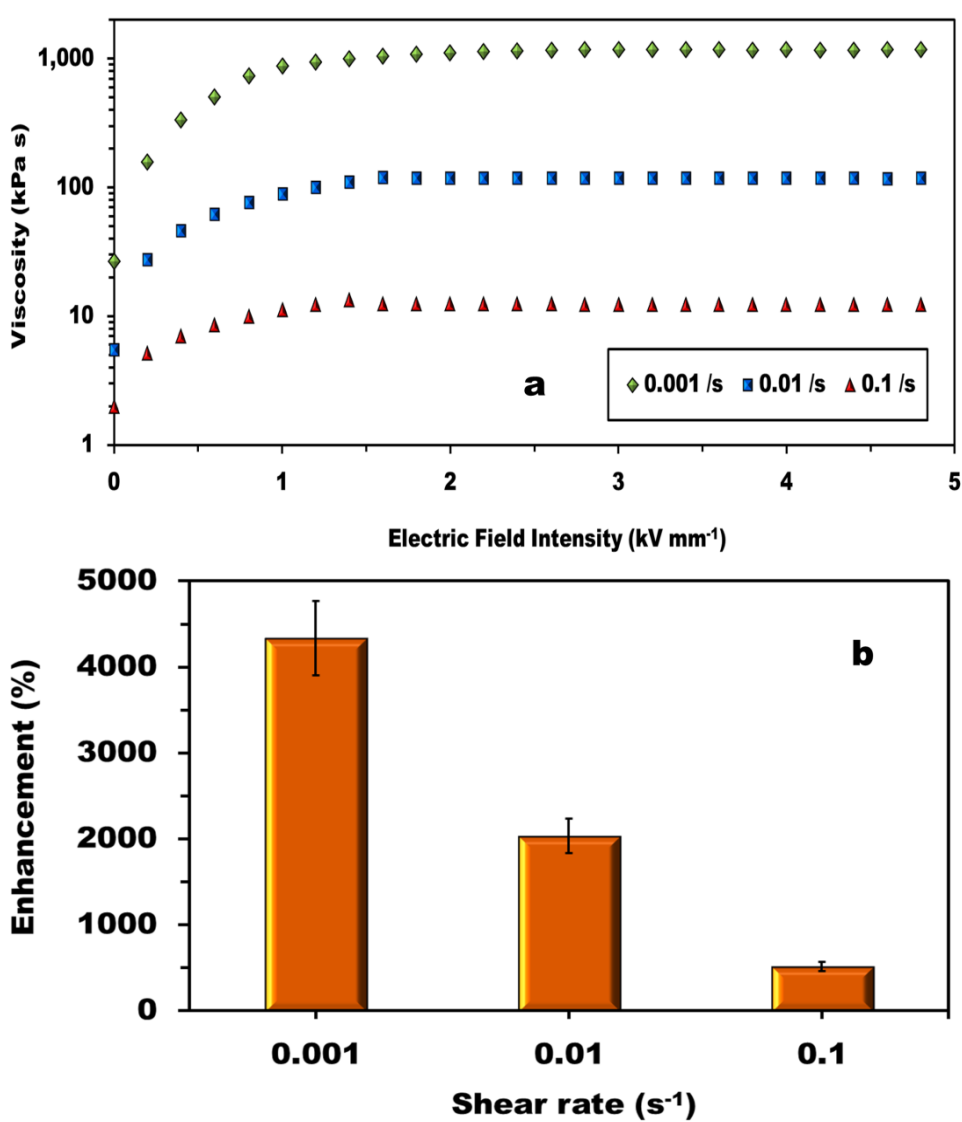

**Figure 4:** Electroviscous response of G0 sample (1.5 wt. % G) **(a)** The response as a function of imposed shear **(b)** The maximum enhancement in viscosity obtained at each shear rate. Shear rates are restricted to lower values due to the inherent nature of ER such that it is appreciably high at low shear values.

The innate advantage of an ER fluid which operates on the mechanism of electrostatic effect is the elimination of leakage current from the system, thereby allowing faster and precise response to electric fields. However, the strong propensity of fibrillation often leads to hampered stability of the colloid when operating at high field intensities. The high chain structure stability leads to complete segregation of the particulate phase from the fluid when the imposed field is removed. In systems containing a third phase which has strong physical affinity towards the fluid phase as well as the particulate phase, such phenomena can be delayed up to very high field intensities. The affinity to the third phase (in this case the trace water content) for the nanoflakes (in this case G) and for the primary fluid (in the present context PEG) ensures that phase separation is prevented even under high field intensities. The affinity for both the phases implies that the water molecules link the G nanoflakes to the gel state PEG by interfacial tension. Therein, the most probable mechanism at play can be elucidated as: the water molecules form interfacial films on the G, which acts as the insulator coat for the conductive nanostructures, leading to efficient formation of fibrils by the electrostatic mechanism, wherein, the water film links the G to the PEG, thereby preventing phase separation by water bridge linkages. Furthermore, the compact texture of the GNGs prevent phase separation and the PEG molecules are expected to form entanglements and chained cages [30] that trap the G nanoflakes, further ensuring high colloidal stability.

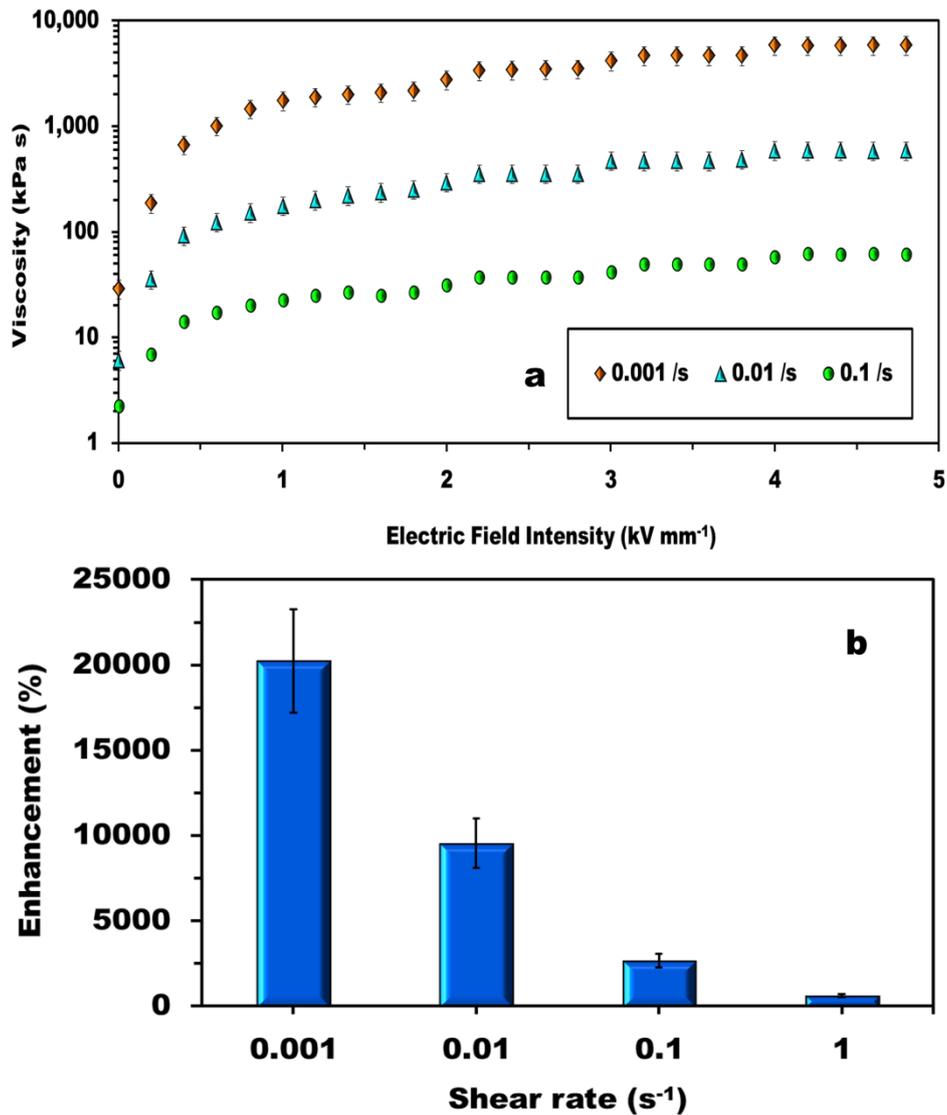

**Figure 5:** Electroviscous response of G3 sample (2 wt. % G) **(a)** The response as a function of imposed shear **(b)** The maximum enhancement in viscosity obtained at each shear rate. This sample exhibits mega electroviscous response.

The high degree of enhancement in viscosity in presence of electric field is a direct consequence of formation of highly stable fibrils, augmented by the high fibril strength due to the relatively larger nanostructure size of G flakes compared to spherical particles commonly utilized in ER fluids. The large surface to volume ratio for nanoflakes, the high delocalized electron density and the compact texture of the gels ensure that full-fledged fibrils are formed at relatively lower field intensities compared to conventional ER fluids. Also the flakes within the fibrils are resilient against the thermal fluctuations since the viscous texture of the gels as well as the meso–scale size of the flakes prevents the same. The viscosity in G0 tends to a saturation plateau beyond certain electric field intensity (Fig. 4 a) whereas G3 tends to a plateau beyond higher field intensities (Fig. 5 a) and such response can be attributed to the difference in G concentrations in the two samples. G3 contains 30 % more G than G0 and this implies that at certain field strengths where all the flakes have contributed to form fibril networks in G0, G3 has excess flakes so as to add to the strength of the existent fibrils. Consequently, viscosity of G3 continues to climb with increasing field intensity beyond the point where G0 attains a plateau. The increment in viscosity under the influence of electric field implies that the sample G3 transits from a gel state to a state mimicking tar at high electric fields and reverts back to the gel state upon withdrawal of the field. The electroviscous effects thus shed light onto the mega electrorheological caliber of the GNGs.

### 3.3. Transient response to electric field

Design of ER fluids often employ comprehensive approaches towards understanding their transient response, since they are often targeted for potential applications in fast and dynamic environment. Thereby it is essential to understand how the electrorheological characteristics of the gels behave under variant transients. Accordingly, the GNGs are subjected to 4 different rates of electric field actuation, viz. 4.8 kVmm$^{-1}$s$^{-1}$, 2.4 kVmm$^{-1}$s$^{-1}$, 1.6 kVmm$^{-1}$s$^{-1}$ and 1 kVmm$^{-1}$s$^{-1}$ while under the influence of constant shear condition of 0.001 s$^{-1}$ and the electroviscous effects therein are observed. Figs. 6 (a) and 7 (a) and (b) illustrate the effect of actuation time on the electroviscous response of the GNGs. As observable, the electroviscous response of the gels is highly dependent on the actuation time of the field. In case of the sample G0 (Fig. 6 (a)), a very rapid actuation of field from zero to its maximum value (at 4.8 kVmm$^{-1}$s$^{-1}$) at very low imposed shear loads leads to reduced magnitude of the maximum attainable viscosity at maximum field. However, interestingly, the system shows no appreciable difference in response for all three other values of actuation rates (at 2.4 kVmm$^{-1}$s$^{-1}$, 1.6 kVmm$^{-1}$s$^{-1}$ and 1 kVmm$^{-1}$s$^{-1}$) and the maximum viscosities attained at maximum field intensities are nearly equal. Furthermore, the trend of increase in viscosity is observed to be independent of the actuation rates, with all the cases exhibiting a non linear trend of increment in viscosity with increasing field intensity, to be followed by a saturation plateau. However, the case is not so in case of higher imposed shear rates. As observed in Fig. 6 (b), the trends in enhancements are markedly different for high and low rates of field actuation. Furthermore, low actuation times are observed to lead to saturation of the electroviscous effect beyond certain field intensity, whereas saturation is never observed at low imposed shear rates. The same is also observed for concentrated systems as G1 in Fig. 7 (a).

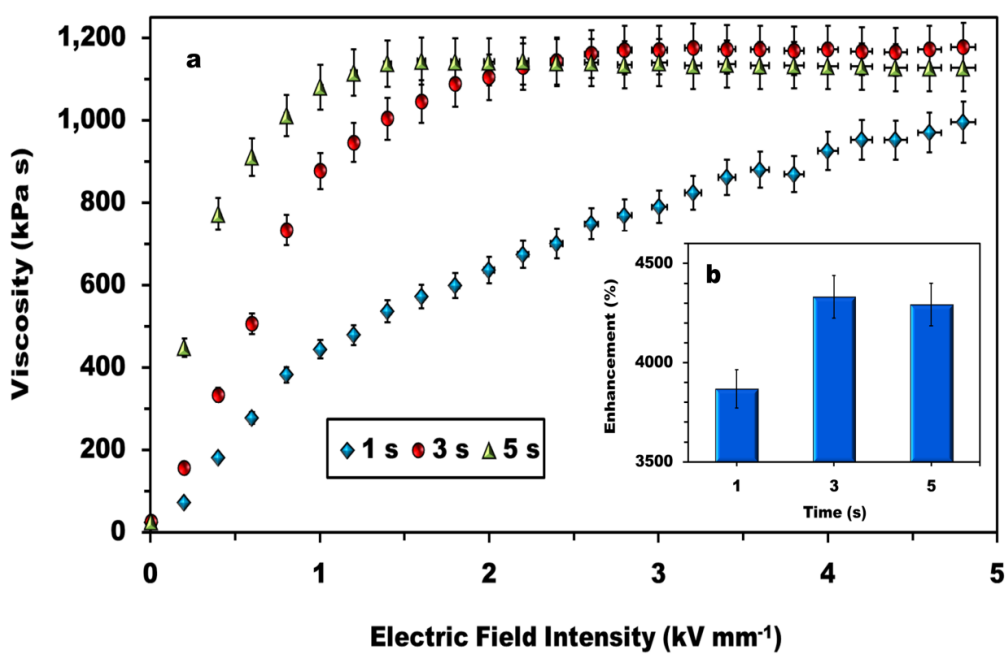

**Figure 6:** Transient electroviscous response of G0 sample. (a) The response as a function of total time of field actuation from zero to maximum intensity has been illustrated for 0.001 s$^{-1}$ imposed shear rate. (b) The maximum enhancement in viscosity obtained at various actuation times at constant shear rate.

Such observations can be explained based on the temporal orientation of the G nanoflakes within the gel due to the simultaneous action of imposed shear and increasing electric field intensity. In case of low shear rates or nearly static conditions, the shearing action of the G flakes with respect to one another is very low, and thereby as the fibrils are formed under the application of a slow actuating electric field, their structural integrity is not compromised by the imposed low shear. Consequently, the gel is able to attain a complete transition to the solid like state and thereby exhibits high viscosity. However, in case of very quick field actuation, the complete G nanoflakes population cannot align itself in the form of fibrils. The fast actuation leads to a smaller fraction of flakes being able to respond to the field and align in to chains, thereby leading to low field induced enhancement in viscosity. At longer actuation times, the whole population is able to attain the complete fibril structure and the viscosity is enhanced. Also, for the present gels, the population seems to respond at its peak for actuation times of 2.4 $kVmm^{-1}s^{-1}$ or more, and hence increment of actuation time beyond 2 seconds does not lead to further enhancement in maximum viscosity obtained.

The phenomenon of attainment of saturation viscosity by the gels under the combined effect of higher shear rates and high field actuation periods can also be explained based on the behavior of the G flakes within the gels. As observable from fig. 7 (a), the sample G1 shows a plateau in case of high shear rates and high actuation time, and the electroviscous effect saturates beyond a certain field intensity. However, in case of low actuation periods, the saturation is not observed and the viscosity rises continually. In the latter case, the short time period of actuation combined with the enhanced shear rate does not induce response fast enough from the G flakes towards fibrillation, leading to gradual chain formation with increasing field intensity, and thereby leading to the gradual rise in viscosity. At higher shear rates, the fibrillation characteristics is hampered due to constant shear induced motion of the flakes against one another, leading to formation of weaker chain networks. However, if the field actuation time is also high, the flakes get enough time to respond to the changing flux and are more potent towards aligning themselves in fibrils. Consequently, the corresponding viscosity is higher than the fast actuation case. However, the shear constraint ensures that the fibrillation can only achieve certain maximum structural integrity at a particular field intensity, beyond which the field has no effect on the fibrillation characteristics. Beyond the critical field intensity, the imposed shear is able to reduce the strength of the fibrils by sliding and frictional action, thereby rendering further increase in solid character null. Since the field is present, the shear cannot disintegrate the fibrils completely and hence due to the competetive interplay, the viscosity attains a plateau.

During field actuation at low shear, the induced shear is unable to disintegrate the fibrils faster than the rate at whch the field induces fibrillation and thus the viscosity increases continually. However, in case the field is applied too fast, the population of flakes forming the fibril at a given instant of time by responding instantaneously to the field is lower than the case of longer actuation time wherein majority of the flakes get enough time to orient themselves along the field lines. As a result, the rise in viscosity is steeper in case of slow field actuation compared to fast field actuation. Fig. 7 (c) illustrates the enhancement in viscosity as a function of field actuation time and shear rate. At shear rate of 0.01 s$^{-1}$ it is observed that the effective enhancement in viscosity is higher in case of the dilute G1 as compared to the concentrated G0 (fig. 6). Furthermore, while the enhancement at fast field actuation is lower than slow field actuation in case of G0, it is nearly equal in the case of G1 and are counter–intuitive observations.

Such an anomaly can be explained based on the principle of lubrication characteristics of Gr/G systems. Gr/G systems are excellent natural lubricants and reduce the frictional force between surfaces by sliding action between the flakes, more so in fluidic systems [31, 32]. The large surface area to volume ratio, accompanied by the flat geometry of the Gr/G flakes leads to high friction reduction characteristics. Higher concentration of G nanoflakes within a system implies decreased proximity between neighboring G nanoflakes within the closely packed fibrils. Consequently, while the fibrils are stronger that those within a low concentration system, the shear induced slipping of the flakes over one another is also higher in a concentrated system. As a result, while the induced viscosity is higher due to presence of a larger flake population density contributing to the fibrillation, the enhancement in viscosity compared to the base viscosity is lower than that in low concentration systems. Due to the additional mechanism of augmented interflake slip creeping into the forefront within such systems, rise in viscosity is hampered at higher shear rates, where the increased relative velocities within the fluid layers induces higher particle slip. The increased shear reduces the friction resistance between the flakes and leads to enhanced slipping, thereby reducing the structural strength of the fibrils against the imposed shear, essentially leading to a lowered maximum enhancement in electroviscosity.

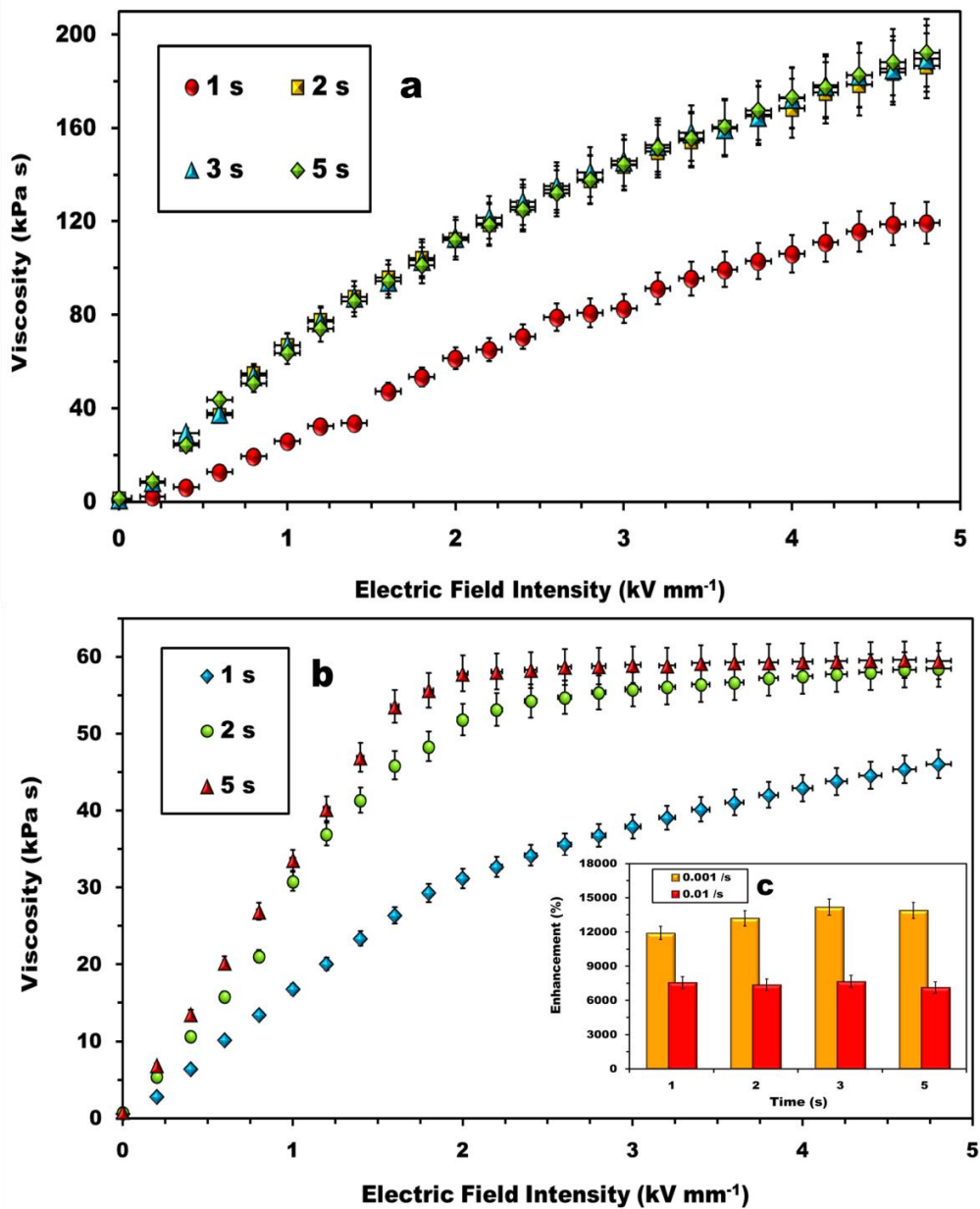

**Figure 7:** Transient electroviscous response of G1 sample. The response as a function of total time of field actuation from null to maximum intensity has been illustrated for (a) 0.001 s$^{-1}$ and (b) 0.01 s$^{-1}$. (c) The maximum enhancement in viscosity obtained at various actuation rates and imposed shear.

### 3.4. Enhanced yield stress

As far as applications of ER fluids are concerned, the two most important aspects that come to the forefront are the electroviscosity and the field induced yield strength. While the former is essential for utility in dynamic systems (where the viscous resistance offered is of prime importance), the latter is important in case of static and sudden loads wherein sudden loading effects are to be absorbed or dissipated without considerable flow. Accordingly, it is of prime importance that the yield strength characteristics and its field response for the GNGs be probed. Figs. 8 (a) and 9 (a) illustrates the response of shear stress to the imposed shear rate at different constant imposed electric field strength for representative GNG samples. As observable, the GNGs behave in general as Bingham plastics with well–defined yield stress whose magnitude enhances with increasing imposed electric field strength. However, it is noteworthy that for some of the samples (especially at high concentration of G), the stress response beyond the yield stress at maximum applied field intensity does not obey the constant deformation characteristics of Bingham plastics. Rather, it has been observed that the stress induced within the sample reduces as the imposed strain increases, hinting at possible disruption of the fibril networks (fig. 8 (a)). However, this has been observed to occur at random in a few cases and could also be due to fatigue within the sample due to continuous experimentation. The appearance of such loss in viscous response appears randomly during the experiments, with a probability that 1 or 2 cases out of 10 continuous runs exhibit such phenomena and therefore it is difficult to conclusively provide a consistent mechanistic theory for the phenomenon. The electric field induced enhancement in yield stress has been observed to be a direct function of the concentration of the G nanoflakes in the system; such that the sample with the highest G concentration (G3) exhibits the highest attainable yield strength. The yield strength of G3 sample has been illustrated in Fig. 10 (a) and the enhancement of yield stress as a function of field intensity illustrated in Fig. 10 (b). It can be observed from the figures that a high degree of enhancement in static stress bearing capacity can be obtained from the GNGs, with enhancements ~ 175 % for the most dilute sample (G2) and ~ 850 % in case of the most concentrated (G3). This in conjunction with the immense electroviscous effect ensures that the GNG samples exhibit large to mega scale electrorheological behavior at very low particulate concnetrations.

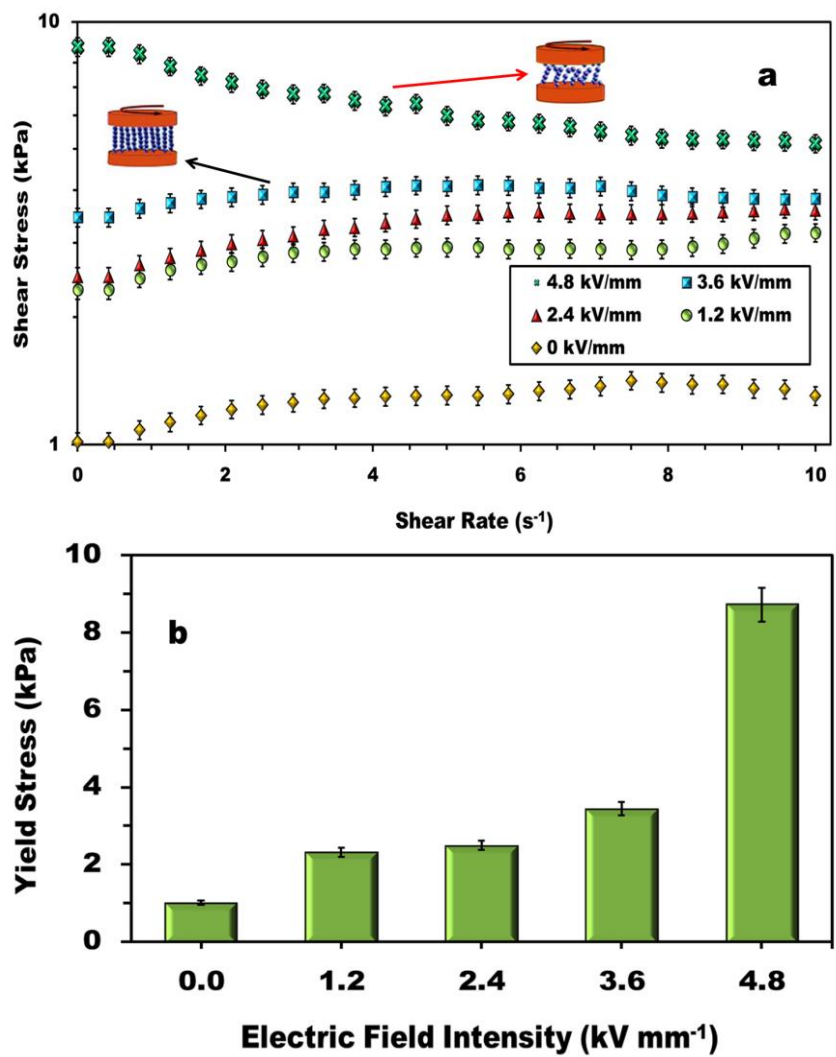

**Figure 8:** (a) Shear stress vs. shear rate plot indicating the electric field induced augmented yield stress for G0. A representative of the possible microstructure failure phenomena for concnetrated systems at high fields has been shown. For moderate field strenghts, the gel shows Bingham plastic behavior. (b) Electric field induced yield stress for G0 as a function of electric field intensity. While at high field the yield stress shoots up, the structural integrity with increase in shear might not always be guaranted.

The highly augmented yield stress bearing capacity adds evidence to the fibrillation by the G nanoflakes, since transition to a near solid state would solely justify yield stresses as high as ~ 13 kPa. Furthermore, the point of interest lies in the fact that such high yield capacity is obtained only at ~2 wt. % G concentration, whereas similar strengths in conventional ER fluids are obtained in general above and beyond ~ 10 wt. % of the dispersed phase concentration. As the electric field intensity is enhanced, a larger portion of the total G flake population contribute to the fibrillation process, leading to the formation of longer fibrils with higher fibril number density per unit volume of the GNGs. Accordingly, the gels experience transition towards a more solid like state and the ability to resist shear deformation enhances. However, beyond certain limit, although certain degree of deformation begins, the structural integrity of the fibril network is not hampered, as evident from the continuous deformation beyond yield point. The augmentation of the yield stress initially follows a linear growth pattern with imposed field strength and slowly transits to a phase of diminishing growth as the field strength grows, as illustrated in Fig. 10 (b). It is intuitive that as the total G population starts contributing to fibrillation with increasing field, the strength of the gel tends to attain maxima at a particular field strength and the yield strength attains a plateau as a function of field intensity.

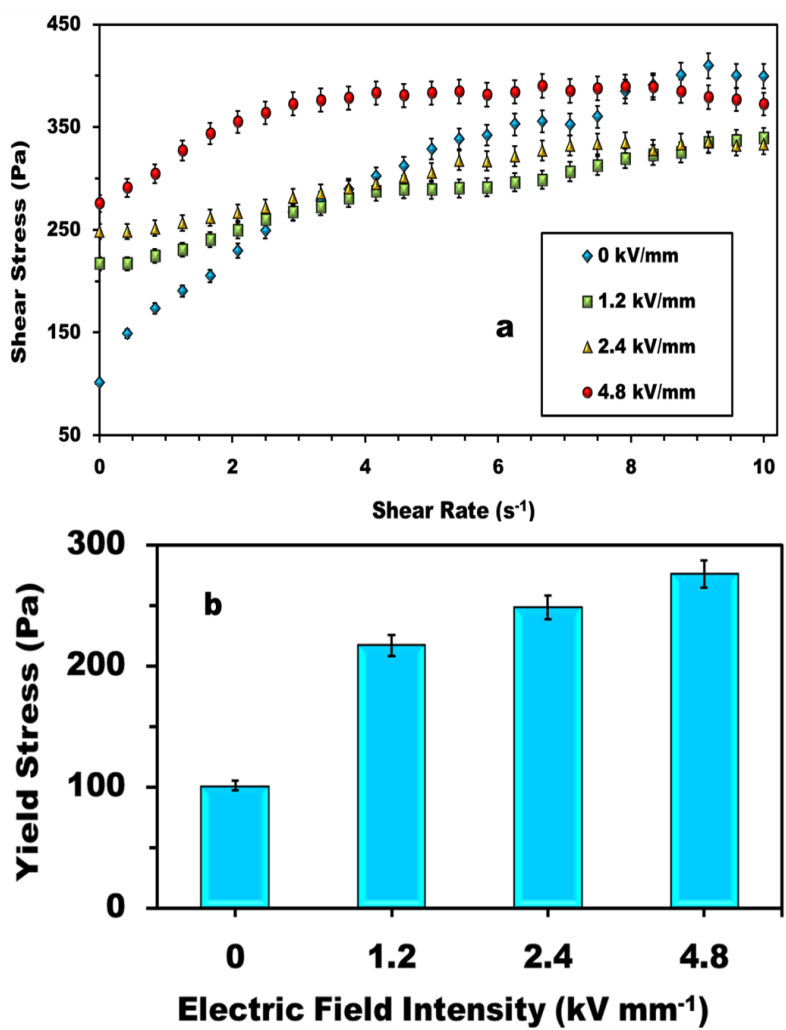

**Figure 9:** (a) Shear stress vs. shear rate plot indicating the electric field induced augmented yield stress for G2. Fairly consistent Bingham plastic behavior is observed for the gels. (b) Electric field induced yield stress for G3 as a function of electric field intensity.

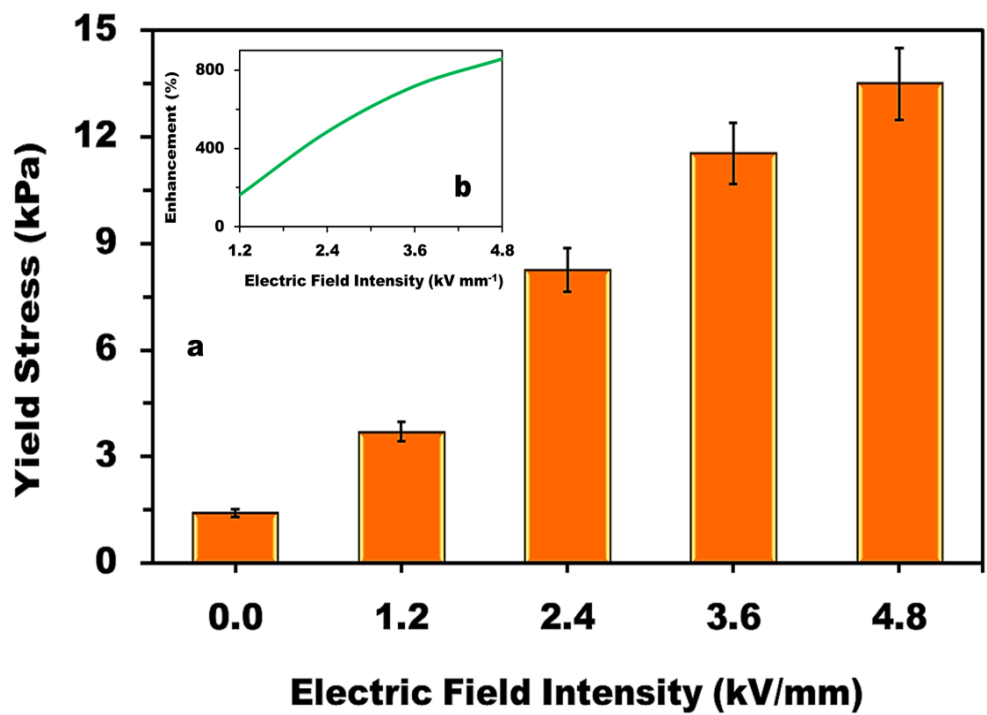

**Figure 10**: (a) Electric field induced yield stress for G3 as a function of electric field intensity. (b) Trend of the enhancement of yield stress as a function of the imposed electric field intensity. The sample G3 (2 wt. % G) exhibits maximum yield capabilities at ~ 13 kPa.

## 4. Conclusions

The present article discusses in depth the large ER effects observed in novel GNGs synthesized utilizing chemically exfoliated G with PEG 400 as the base medium. Detailed evidence and discussions on the fibrillation characteristics of the G nanoflakes under the influence of external electric fields have been presented, thus establishing the possibility of ER response from G based systems even at low field strengths. The high polarizability of the flakes caused by the presence of the dense delocalized π electron cloud in G has been theorized to be the backbone of the electro–fibrillation phenomena. This enables the G nanoflakes to respond to the external field and align themselves along the field lines, forming fibrils. The GNGs have been reported to exhibit electrorheological effects in the range of high to large, depending upon the concentration of G flakes within the system and with respect to the imposed system parameters such as shear, electric field intensity and field actuation period. The viscosity has been observed to enhance by many orders of magnitude upon application of strong electric fields and based on representative magnitudes of viscosity, the transition can be compared to that from a gel to tar/pitch state ( enhancements of the order of ~ 20000 % has been recorded). The gels have been theorized to enhance viscosity by a mechanism comprising of the dual nature of interfacial tension and electrostatic theories of conventional ER materials. The transient response of the GNGs to field actuation time has also been discussed and the effect can be tuned for applications in various time reliant systems. The yield strength of the gels, the most important characteristic of an ER medium, has been observed to be greatly enhanced by the presence of an external electric field. The magnitudes of yield stresses obtained from the dilute GNGs are in general possible only for very high concentrations of traditional nano ER systems. The present gels and their huge ER response show potential and promise as futuristic smart damping and vibration control fluids, as braking, actuation and clutching fluids in locomotives and automotive systems and assembly lines, in robotic flexible joint manipulation, as fluidic sensors and actuators in MEMS/ NEMS and so forth. An increase in viscosity from ~ 30 kPas to ~ 5000 kPas at an electric field intensity of ~ 4.8 kV/mm and shear rate of 0.001 $s^{-1}$ for 2 wt. % G concentration and high viscosity even at low dynamicity conditions ~ 10 $s^{-1}$ implies that such GNGs can be effectively used as smart damping, actuating or shock absorber medium in near static systems, such as heavy machinery foundations, aircraft landing gear suspension systems, locomotive suspensions, seismic damping for building columns etc. The high viscosity of the gels even in the absence of electric fields and the biocompatibility of PEG makes them possible potential candidates for shock absorption in biomedical implants and prostheses.

## Acknowledgement

The authors would like to thank Dr. T Nandi, Scientist 'F' and Head, Fuel and Lubricants Division of the Defence Materials and Stores Research and Development Establishment (DMSRDE) Kanpur (a DRDO laboratory) for the rheometer facility. PD would also like to thank IIT Madras and IIT Ropar for partial financial support towards the present work.